\documentclass[11pt,letterpaper,]{article}
\usepackage[T1]{fontenc}
\usepackage[utf8]{inputenc}
\usepackage{lmodern}
\usepackage{upgreek}
\usepackage{textcomp}
\usepackage{amsmath,amsfonts,amssymb,amsthm,mathrsfs}
\usepackage{latexsym}
\usepackage{authblk}
\usepackage{graphicx}
\usepackage[english]{babel}
\usepackage{indentfirst,setspace,fancyhdr}
\usepackage{xcolor}
\usepackage{soul}
\usepackage{cite}
\usepackage[normalem]{ulem}
\usepackage[outercaption]{sidecap}
\usepackage[top=2cm, bottom=2cm, left=2cm, right=2cm]{geometry}

\usepackage{sectsty}
\allsectionsfont{\sffamily}

\makeatletter
\renewcommand\@biblabel[1]{#1.} 
\makeatother

\usepackage{setspace}
  
\emergencystretch=\maxdimen
\begin{document}
\doublespacing 
\selectlanguage{english}

\noindent {{\bf \sffamily  Title:}  \\
Nuclear compensatory evolution driven by mito-nuclear incompatibilities }

\vspace{0.7cm}

\noindent {\bf \sffamily Authors:}\\
{Débora Princepe}$^{1,2,*}$,
{Marcus A. M. de Aguiar}$^1$

\vspace{0.7cm}
\noindent {\bf \sffamily  Affiliation:} \\
1. {Instituto de F\'{i}sica Gleb Wataghin, Universidade Estadual de Campinas (UNICAMP), Campinas, Brasil - 13083-859}\\
2. {Quantitative Life Sciences Section, The Abdus Salam International Centre for Theoretical Physics (ICTP), Trieste, Italy - 34151 (current institution)}

\vspace{0.7cm}
\noindent {\bf \sffamily  Corresponding author:} \\
Débora Princepe\\
c/o QLS Section - Strada Costiera, 11 - 34151 - Trieste, Italy \\
phone: +39 040 2240139 \\
e-mail: dprincep@ictp.it 

\vspace{0.7cm}

\noindent {\bf \sffamily  Classification: }\\
BIOLOGICAL SCIENCES - Evolution

\vspace{0.7cm}

\noindent {\bf \sffamily  Keywords: }\\
Mito-nuclear co-evolution; Nuclear compensation; mtDNA introgression; Mitochondrial mutation rate

\newpage
\begin{center}
   {\bf \sffamily \LARGE Nuclear compensatory evolution driven\\ by mito-nuclear incompatibilities}
\end{center}

\section*{Abstract}

Mitochondrial function relies on the coordinated expression of mitochondrial and nuclear genes, exhibiting remarkable resilience regardless the susceptibility of mitochondrial DNA (mtDNA) to accumulate harmful mutations. A suggested mechanism for preserving this mito-nuclear compatibility is the nuclear compensation, where deleterious mitochondrial alleles drive compensatory changes in nuclear genes. However, prevalence and conditioning factors for this phenomenon remain debated, with empirical evidence supporting and refuting its existence. Here, we investigate how mito-nuclear incompatibilities impact nuclear and mitochondrial substitutions in a model for species radiation under selection for mito-nuclear compatibility, similar to the process of mtDNA introgression. Mating eligibility relies on genetic (nuclear DNA) and spatial proximity, with populations evolving from partially compatible mito-nuclear states. Mutations do not confer advantages nor disadvantages, with no optimal nuclear or mitochondrial types, but individual fitness decreases with increasing incompatibilities, driving the demand for mito-nuclear genetic coordination. We find that selection consistently promotes compensation on incompatible nuclear genes, resulting in more substitutions than compatible or non-interacting genes. Surprisingly, low mitochondrial mutation rates favor compensation, as do increased selective pressure or a higher number of mismatches. High mitochondrial mutation rates boost substitutions in initially compatible nuclear genes, relaxing the selection against mito-nuclear incompatibilities and mirroring the compensatory evolution. Moreover, the presence of incompatibilities accelerates species radiation, but richness at equilibrium is not directly correlated with substitutions' response, revealing the complex dynamics triggered by mitochondrial introgression and mito-nuclear co-evolution.

\section*{Significance Statement}
Addressing how nuclear and mitochondrial genomes co-evolve is imperative for understanding organismal resilience and adaptation, given the relevance of their compatibility for cellular functioning. This study explores how mismatches between interacting nuclear and mitochondrial genes impact substitutions, shedding light on nuclear compensation -- a process where deleterious mitochondrial mutations induce compensatory changes in corresponding nuclear genes. Using a model for speciation with selection for mito-nuclear compatibility, we find that low mitochondrial mutation rates, high selective pressure, and increased mismatches facilitate nuclear compensation. Notably, substitutions can overcome neutrality in initially compatible nuclear genes under high mitochondrial mutation rates, thus relaxing the purifying selection. These findings contribute to understanding population resilience despite genetic incompatibilities and the connections between mito-nuclear co-evolution and substitution rates.

\section*{Introduction}
Mitochondrial DNA (mtDNA) is a remarkable genetic component for maintaining the myriad of mitochondrial functions in cellular metabolism, distinguished by its small size, rapid evolutionary rate, and general absence of recombination associated with uniparental inheritance. These characteristics render mtDNA susceptible to the irreversible accumulation of deleterious alleles, known as ``Muller's ratchet'' \cite{Howe2008, Rand2001, Hill2020}. Evidence of multi-level purifying selection, spanning from cellular organization to individual and population levels, provides the most plausible explanation for the non-observation of mutational erosion \cite{Stewart2008, Camus2023,Hill2020b}. However, harmful mutations may evade the purifying mechanism when linked to advantageous genes, indicating that our understanding of why mitochondrial function does not decline over time is still limited \cite{Hill2020}.

A manifestation of selection arises from the crucial interaction between proteins encoded by nuclear and mitochondrial DNAs for cellular respiration. These genomes must co-evolve to maintain mito-nuclear genetic compatibility, ultimately impacting an organism's phenotype \cite{Bettinazzi2023, Hill2019b, Rank2020, Healy2020}. Complementary co-evolution occurs when a neutral or beneficial mutation in one genome enables further changes in the corresponding genes of the other, resulting in increased fitness. Conversely, the proposed mechanism of ``nuclear compensation'' suggests that mitochondrial deleterious alleles drive compensatory changes in nuclear genes to preserve mito-nuclear co-adaptation and functional mitochondria \cite{Hill2019, Zwonitzer2023, Havird2016}. Experimental test of this hypothesis requires distinguishing between positive and relaxed purifying selection \cite{Weaver2022, Havird2017, Zwonitzer2023}, as different scenarios of mito-nuclear co-evolution depend on how mitochondrial mutations impact fitness relative to the nDNA background \cite{Sloan2017}. Moreover, interpretation of the selective forces at play depends on hypothesis formulation and testing tools employed \cite{Zwonitzer2023, Weaver2022}. Considering all this, empirical evidence both supports \cite{Barreto2013, Barreto2018, Osada2012} and refutes \cite{Zhang2013, Weaver2022, Piccinini2021} nuclear compensation, leaving the debate unresolved regarding the ubiquity of such a mechanism and the factors that condition its occurrence.

\begin{figure}[h]
\centering
\includegraphics[width=0.5\linewidth]{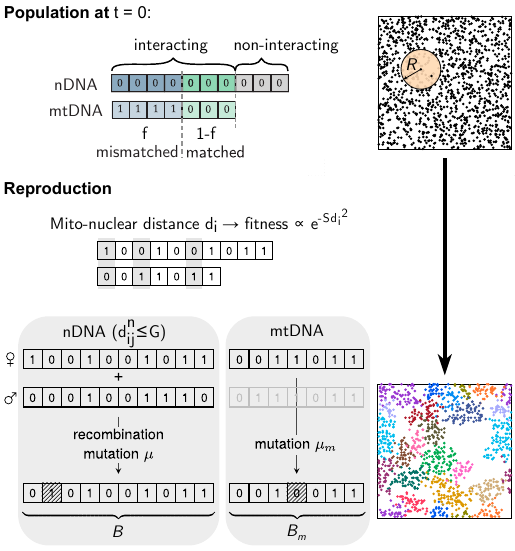}
\caption{Model key components. The upper section illustrates the initial population spatially distributed in a homogeneous area. Each individual carries two sequences representing its nuclear and mitochondrial DNAs. All nuclear genes initiate on zero. A fraction $f$ of incompatible mitochondrial genes are set on one (mismatched, in blue) and the remaining set on zero (matched, in green). During reproduction (lower section), individuals mate within a spatial neighborhood of radius $R$. Individual's fitness determines the mating probability as a function of its mito-nuclear distance $d_i$ (total incompatibilities, highlighted sites), following $e^{-S d^2_i}$, where $S$ denotes the selection strength. Mating partners must have nDNA's genetic distance $d^n_{ij}$ smaller or equal to a threshold value $G$. Offspring is produced for the next generation with free recombination of the parents' nDNA with a probability of mutation $\mu$ per gene and receive mtDNA from the mother, with a mutation probability $\mu_m$. Over time, species may arise, as shown by different colors in the spatial plot at the bottom right.}
\label{fig:model}
\end{figure}

Here, we study the impact of mito-nuclear incompatibilities on nuclear and mitochondrial substitutions in a model simulating species radiation. We aim to distinguish signatures of mito-nuclear co-evolution across various scenarios, particularly identifying conditions that promote compensatory mutations in nuclear DNA (nDNA) to restore compatibility with the corresponding mtDNA part. To this end, we employ agent-based simulations where genetically explicit populations evolve from partially compatible mtDNA and nDNA. All individuals have a fraction $f$ of mitochondrial genes incompatible with their respective interacting nDNA sites (first nDNA/mtDNA genes in the upper section of Fig. \ref{fig:model}). We assume no optimal nuclear or mitochondrial types, but their compatibility influences individual fecundity: individual fitness decreases with mito-nuclear incompatibilities, exerting selective pressure towards matching genetic sequences.  

Nuclear DNA also governs mating compatibility, meaning individuals with significantly different nDNA cannot reproduce. Reproduction is further constrained to spatially proximate individuals, leading to species emergence under the influence of mito-nuclear selection. We vary the strength of selection ($S$), the initial fraction of incompatible genes ($f$), and the mitochondrial mutation rate ($\mu_m$), and track how substitutions vary over time for the nuclear and mitochondrial genomes, distinguishing sections that start as incompatible (\textit{mismatched}) or compatible (\textit{matched}), and those unaffected by selection (\textit{non-interacting}, Fig. \ref{fig:model}).

In this simplified scheme, the described process resembles the dynamics after mtDNA introgression, where foreign mtDNA invades the population and potentially introduces incompatibilities with the native nuclear genome \cite{Morales2018}. These transfers commonly occur when previously isolated populations come into secondary contact followed by hybridization and back-crossing \cite{Darling2011, Shu2022, Boratynski2014}. Two scenarios can explain the observation of mtDNA introgression: it can occur adaptively, providing a fitness advantage in a changing or novel environment, or by replacing a mitochondrial genome with a high mutational load \cite{Sloan2017, Hill2019c, Nikelski2022}. Indeed, there are numerous cases where mitochondrial introgression seems to occur without noticeable migration of nuclear genes, mainly due to sex-biased phenomena or selective sweeps \cite{Boratynski2014, Sloan2017, Toews2012}. We assume that any of these drivers are at play and explore the outcomes to address the seeming paradox between mitochondrial introgression and selection for mito-nuclear compatibility \cite{Burton2022, Hill2019c}. Our study can thus offer insights into conditional factors for nuclear compensation and whether substitution rates might indicate the evolutionary stage of the population.

\section*{Results}

\subsection*{Model} 
We investigated the impact of mito-nuclear incompatibilities during the radiation of spatial populations, employing an approach previously explored in \cite{Princepe2021,Princepe2022a}. The model incorporates sexual reproduction with separated sexes, dispersal, mutation, and genetic drift as processes that change allele frequencies. Together with conspecific mating restricted to nearby individuals, species emerge through isolation by distance \cite{aguiar_global_2009}. Our study involves individual-based simulations where initially identical individuals are characterized by haploid biallelic sequences of lengths $B$ and $B_m$, with $B>B_m$, representing their nuclear and mitochondrial DNAs, respectively (Fig. \ref{fig:model}; Materials and Methods). The mito-nuclear genetic interaction is modeled as a locus-by-locus coupling between mtDNA and the first $B_m$ genes of the nuclear DNA, while the remaining $(B-B_m)$ sites, designated ``non-interacting genes'', do not participate in the interaction. Pairs of interacting loci that carry different allele values are considered incompatible and contribute to the mito-nuclear distance $d_i$, which in turn affects an individual's fecundity. This scheme is a phenomenological approximation of the required, intricate biochemical and structural compatibility between proteins encoded by both genomes for cellular respiration \cite{Bettinazzi2023, Zhang2013}.

Populations evolve with non-overlapping generations and in a homogeneous environment, maintaining a constant total number of individuals. During the reproduction phase, each individual has a probability of randomly choosing a mating partner within a spatial range of radius $R$. Genetic compatibility depends on the Hamming distance between their nuclear DNAs ($d^n_{ij}$), which must be below the threshold $G$ for successful mating. Offspring's nuclear genome forms through parental free recombination with a mutation probability of $\mu$ per site, while mtDNA is inherited from the mother with a mutation probability $\mu_m$ per locus (Fig. \ref{fig:model}). It's worth noting that mutations can reverse in our simulations. Species emerge as clusters of genetically compatible individuals with potential gene flow, delimited by the genetic mating restriction $G$, with no gene flow between clusters. 

In the absence of mito-nuclear selection, all individuals have an equal probability of reproduction, and nDNA and mtDNA evolve independently. However, under selection for mito-nuclear compatibility, an individual's fitness is determined by their mito-nuclear distance, expressed as $e^{-Sd_i^2}$, where $S$ represents the strength of selection -- high values indicate strong selection. Individuals with higher fitness, normalized within the mating range, have a higher probability of reproduction, both as focal individual and drawn partner. We evaluated the impact of partially incompatible genomes, where the entire population initially possesses a fraction $f$ of incompatible mtDNA genes, under various selection regimes and mitochondrial mutational rates. 
We sought signs of relaxed selection and compensatory nuclear evolution by tracking all substitutions, aiming to understand the specific conditions under which these phenomena may occur. This simplified description does not intend to replicate biological systems, but rather to capture the fundamental processes underlying species evolution under selection for mito-nuclear coordination, described as a stabilizing mechanism. 

The detection of selection often relies on the ratio between non-synonymous to synonymous nucleotide substitution rates \cite{Weaver2022, Zwonitzer2023}. Purifying selection removes mutations, while positive selection favors their fixation, changing the fixation frequencies relative to the baseline evolution rate \cite{Lynch2023}. In this study, we use the term ``substitutions'' to denote mutations that become fixed in the population, changing from an initial state of 0 to a final state of 1 or vice versa \cite{Rand2001}. It is important to note that mutations in our model are neutral, meaning they have no direct impact on individual fitness and are equivalent among genes, conferring no evolutionary advantage or disadvantage. However, any mutation on matched genes will cause fitness loss if not followed by a change in the counterpart. Consequently, the dynamics on incompatible gene pairs resemble compensatory rather than complimentary co-evolution \cite{Hill2020}, with stabilizing selection acting towards minimizing mito-nuclear mismatches. We do not distinguish between synonymous and non-synonymous substitutions; in a simplified manner, all substitutions in our model can be considered non-synonymous.

\subsection*{Mitochondrial Mutation Rates}

\begin{figure}[!b]
\centering
\includegraphics[width=0.5\linewidth]{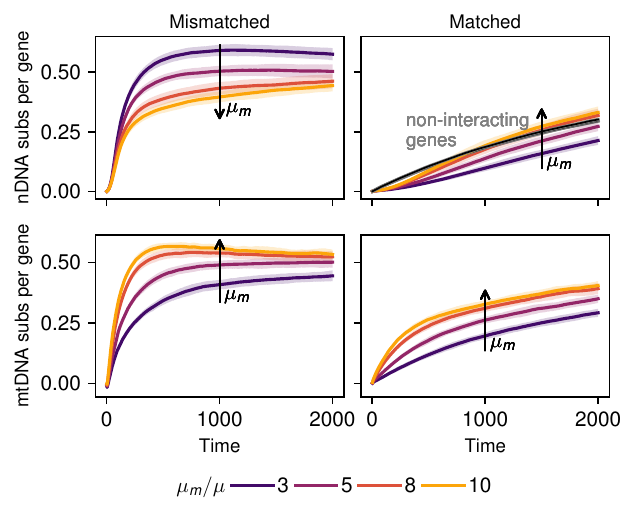}
\caption{Effect of varying mitochondrial mutation rate on substitutions. Solid lines represent the ensemble mean of 50 independent realizations, with shaded areas indicating one standard deviation. We fixed the initial fraction of incompatible genes $f=0.5$ and strong selection ($S=90$).
On nuclear sites, the impact of $\mu_m$ depended on the initial state: for initially incompatible genes, increasing $\mu_m$ reduced the number of substitutions per gene (top left), while for compatible sites, higher $\mu_m$ increased it (top right). Substitutions at non-interacting nuclear genes were independent of $\mu_m$ (depicted in black). For mtDNA, increasing $\mu_m$ led to more substitutions in both sections (bottom), occurring more on mismatched sites.}
\label{fig:Fig2}
\end{figure}

Figure \ref{fig:Fig2} shows the impact of varying mitochondrial mutation rate on nuclear and mitochondrial substitutions in genes initially compatible or incompatible. Increasing $\mu_m$ consistently increased the number of fixed mutations in mitochondrial genes in mismatched and matched sites (bottom panels in Fig. \ref{fig:Fig2}), with mismatched sites exhibiting a higher number of substitutions compared to matched ones in the same conditions (compare curves of the same color on left and right). In contrast, increasing $\mu_m$ led to decreased substitutions in nuclear genes when initially incompatible (top left), while increasing it for sites initially compatible (top right). Nuclear mismatched genes also exhibited a higher number of substitutions compared to matched ones in the same conditions (curves of the same color on left and right). Non-interacting nDNA genes remained unaffected.

\begin{figure*}[ht]
\centering
\includegraphics[width=\linewidth]{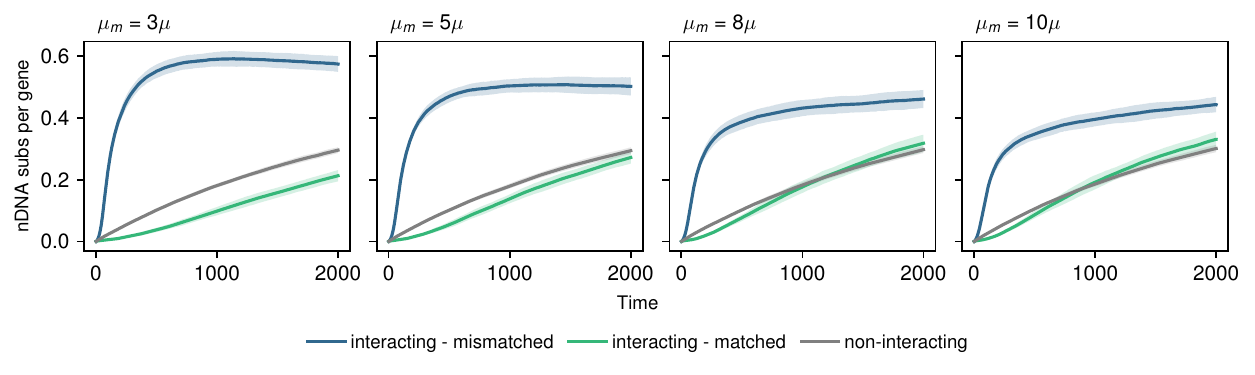}
\caption{Nuclear substitutions under varying mitochondrial mutation rate (same data from Fig. \ref{fig:Fig2}). Interacting genes with initial incompatible states exhibited a notably higher substitution rate compared to compatible or non-interacting sites, attributed to compensatory evolution. In contrast, compatible sites generally experienced a lower substitution rate due to stabilizing (purifying) selection. However, with $\mu_m/\mu \geq 8$, initially compatible sites presented a higher rate than non-interacting sites, evincing relaxed selection.}
\label{fig:Fig3}
\end{figure*}

In Figure \ref{fig:Fig3}, we examine the trade-off between mitochondrial mutation rates and nuclear substitutions across these scenarios. Non-interacting genes were a reference for neutrality within the nuclear genome. We observed that mito-nuclear selection consistently increased substitutions beyond neutrality in initially incompatible nDNA genes, indicating a compensatory effect. A lower $\mu_m$ was advantageous for this effect because, in our simulations, faster mtDNA mutation rates facilitated the reversal of incompatibilities more readily. Conversely, selection had a purifying action on initially compatible nDNA genes for lower $\mu_m/\mu$ ratios, reducing substitutions. However, as $\mu_m$ increases ($\mu_m/\mu \geq 8$), substitutions occurred beyond neutrality even for those genes. In summary, a lower $\mu_m$ facilitated the nuclear compensatory effect on incompatibilities, whereas higher $\mu_m$ increased substitutions beyond neutrality in initially compatible genes due to the rapid accumulation of mutations in the mtDNA.

\begin{figure}[ht]
\centering
\includegraphics[width=0.5\linewidth]{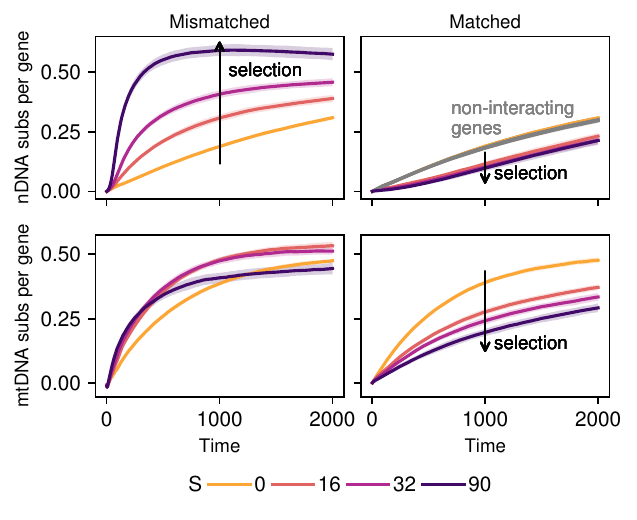}
\caption{Increasing selection strength and substitutions. We run simulations (50 replicates) with an initial fraction of incompatible genes set at $f=0.5$ and a constant $\mu_m/\mu=3$ ratio. Solid lines depict the average values, with shaded areas indicating one standard deviation. 
Selection for mito-nuclear compatibility exhibited a purifying effect on initially compatible sites (right column, `Matched') in both nuclear (nDNA) and mitochondrial (mtDNA) genes, reducing the number of substitutions compared to the absence of selection ($S=0$). In incompatible nuclear genes (top left), increasing selection strength directly promoted nuclear compensation. In the corresponding incompatible mitochondrial genes (bottom left), the effect was non-linear: weak selection ($S=16$ and $S=32$) increased substitutions, but strong selection ($S=90$) suppressed them below neutrality. }
\label{fig:Fig4}
\end{figure}

\subsection*{Strength of Selection}
The effect of selection strength on nuclear substitutions differed from that of the mitochondrial mutation rate. As depicted in the top panel of Figure \ref{fig:Fig4}, increasing selection caused an increase in the number of substitutions on incompatible sites but a reduction in the compatible ones. Compensation was more intense the stronger the selection. For mitochondrial genes, substitutions at mismatched sites (bottom left in Fig. \ref{fig:Fig4}) exceeded neutrality (referenced as $S=0$) under weak and intermediate selection ($S=16$ and $S=32$). However, under strong selection ($S=90$), substitutions initially accelerated before reaching a plateau below neutrality. In matched mtDNA genes, the purifying effect was positively correlated with the strength of selection.

\subsection*{Initial Fraction of Incompatibilities}
Finally, we examined how substitutions were affected by the number of incompatible pairs of genes. Despite loci being independent in our model, the number of substitutions per gene in mismatched nDNA sites increased with $f$ while decreasing in the corresponding mtDNA genes. This observation provides direct evidence that incompatibilities promote nuclear compensation. Meanwhile, matched genes presented similar dynamic responses, roughly independent of the number of incompatibilities.

\begin{figure}[ht]
\centering
\includegraphics[width=0.5\linewidth]{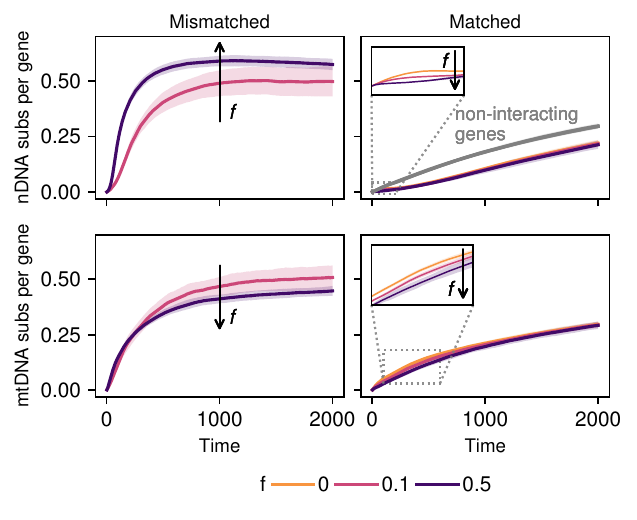}
\caption{More incompatibilities amplify nuclear compensation. Solid lines represent the average across 50 simulations carried out under strong selection ($S=90$) and a fixed $\mu_m/\mu=3$ ratio. Shaded regions denote one standard deviation. Substitutions at incompatible genes were sensitive to $f$ in both genomes (left): increasing the number of incompatible genes boosted nuclear compensation in mismatched sites while suppressing mitochondrial substitutions in the corresponding mtDNA genes. Compatible sites demonstrated greater robustness (right); insets highlight that substitutions in both genomes were slightly different at the beginning of the dynamics but reached similar behavior over time.}
\label{fig:Fig5}
\end{figure}

\subsection*{Mito-nuclear Incompatibilities and Speciation}
\begin{figure}[ht]
\centering
\includegraphics[width=\linewidth]{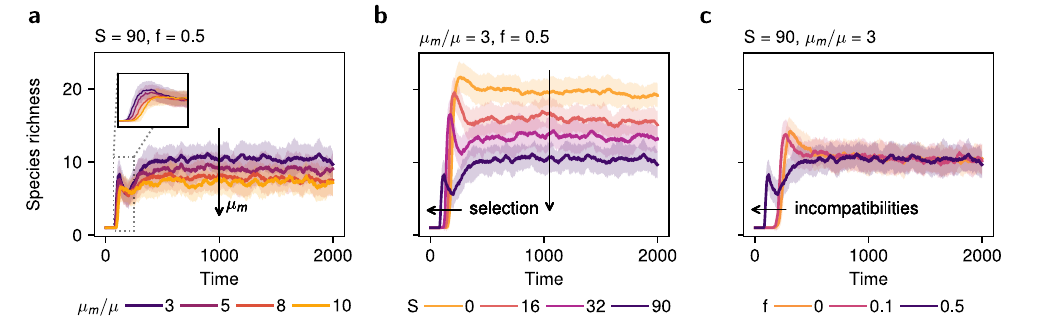}
\caption{Impact on speciation. Solid lines depict the average across 50 simulations, with two fixed parameters for each plot: selection strength $S=90$, fraction of initially incompatible genes $f=0.5$, and the ratio between mutation rates $\mu_m/\mu=3$. The purple curve coinciding in all graphs represents when these three parameters are fixed. (a) Increasing the mitochondrial mutation rate reduced the number of species at equilibrium without significantly affecting the radiation time (inset). (b) Richness also decreased with increasing selection pressure, accelerating radiation in the presence of incompatibilities. (c) An increase in the number of mito-nuclear mismatches led to earlier radiation.}
\label{fig:Fig6}
\end{figure}

The presence of incompatibilities ultimately influenced the radiation process. Increasing $\mu_m$ reduced the number of species in equilibrium, but had a negligible effect on the radiation time (Fig. \ref{fig:Fig6}a). This finding contradicts a previous suggestion that elevated mtDNA substitution rates compared to the nuclear genome might enhance the potential role of mito-nuclear incompatibilities in speciation \cite{Burton2022}. Similarly, stronger selection reduced richness (Fig. \ref{fig:Fig6}b). Thus, while variation in mitochondrial mutation rate and selection intensity had contrasting effects on nuclear substitutions (compare Fig. \ref{fig:Fig2} and Fig. \ref{fig:Fig4}), they yielded similar outcomes regarding richness, indicating that the dynamics of specific segments of the nDNA alone may be insufficient for predicting species evolution.  Regarding the initial level of incompatibilities, the number of species at equilibrium remained unchanged with varying $f$.

Figure \ref{fig:Fig6}c shows that the presence of incompatibilities accelerated radiation when all other parameters were constant. A similar effect was observed with increasing selection strength, particularly pronounced with strong selection (Fig. \ref{fig:Fig6}b). Notably, prior research has shown that, in this model, selection delays the radiation process when starting from a population with fully compatible sets of genomes \cite{Princepe2021}. However, when initiated with incompatible mitochondrial and nuclear DNAs, radiation under selection occurred earlier (Fig. \ref{fig:Fig6}b).

\section*{Discussion}

We investigated the impact of mito-nuclear incompatibilities on nuclear and mitochondrial substitutions using a simplified model for species radiation, simulating the dynamics following mtDNA introgression under selection to minimize mito-nuclear mismatches. 
Our study qualitatively demonstrated nuclear compensation in incompatible nDNA genes, which exhibited consistently more substitutions than compatible and non-interacting genes. Parameters' dependence on mismatched nDNA genes opposed the trends observed in corresponding mtDNA genes and compatible pairs, resembling another signature of compensation.

Mito-nuclear selection was based on a stabilizing mechanism, driving genomes towards minimal discordance ($d_i \rightarrow 0$), but the effect on substitutions depended on the initial genetic compatibility. Selection drove substitutions above neutrality in incompatible nuclear genes, indicating a positive response, i.e., nuclear compensation. In contrast, compatible genes exhibited lower substitution rates, reflecting a purifying character. Yet, under a high $\mu_m$, they exhibited rates surpassing neutrality while decreasing substitutions in incompatible genes (Fig. \ref{fig:Fig3}), suggesting purifying selection becomes inefficient (relaxed) as it cannot keep up with the rate of changes. The observation of a signature resembling nuclear compensation prompted by purifying selection highlights the complex interplay between selection and genetic adaptation conditioned by the mitochondrial mutation rate \cite{Havird2016}, challenging the interpretation of high dN/dS ratios as evidence of positive selection \cite{Zwonitzer2023,Weaver2022}.

Increasing $\mu_m$ was disadvantageous to the compensatory effect because faster mtDNA mutation rates facilitated the rectification of incompatibilities, owing to the reversibility of mutations in our simulations. This surprising result raises questions about the real-life parallelism of the model, contradicting the general expectation that high rates of mtDNA mutation select for compensatory changes in the nuclear genome \cite{Havird2016, Morales2018}. However, this might hold in the context of introgression: highly mutable mtDNA invading an incompatible nDNA background can potentially facilitate mutation for the rescue of compatibility more readily than the nDNA reaching its evolutionary pace.

The response above neutrality in initially compatible genes under high $\mu_m$ suggests that signatures resembling nuclear compensation may arise even without mtDNA introgression or pre-existing incompatibilities \cite{Weaver2022}, noting that substitutions accumulated more slowly in compatible than in incompatible genes (Fig. \ref{fig:Fig3}). A correlation between fast-evolving mtDNA and increased evolution of interacting nuclear genes was demonstrated in angiosperms \cite{Havird2015, Havird2017}, but it was not supported in mammals \cite{Weaver2022}, among other examples. While differences in metabolism or genetic structure between these groups may explain the contrasting observations, our results suggest that the dynamics may also depend on previous mito-nuclear states, motivating further research that could involve exploiting the variability in rates of mtDNA evolution across eukaryotes \cite{Havird2015}.

The observed trends in nuclear substitutions due to parameter variation did not correlate with resulting species richness: although increasing selection and mitochondrial mutation rate had opposing effects on nDNA substitutions, they had the same impact on richness, suggesting that the dynamics of nDNA segments alone may not fully explain species evolution. Furthermore, incompatibilities accelerated radiation, as evidenced directly by increasing $f$ (Fig. \ref{fig:Fig6}c). Increasing selection provided indirect evidence: radiation occurred earlier when incompatibilities were present ($f=0.5$, Fig. \ref{fig:Fig6}b) but was delayed when starting from matched sequences ($f=0$ demonstrated in \cite{Princepe2022a, Princepe2021}). 

The consistent equilibrium richness, independent of the initial number of incompatibilities (Fig. \ref{fig:Fig6}c), indicates that additional system conditions may influence population resilience. Here, we posit a connection to the constant resource availability represented by a fixed number of individuals. These results reconcile mito-nuclear introgression and speciation, illustrating the maintenance of mito-nuclear co-evolution despite the expected selection against mtDNA introgression through simple mechanisms \cite{Nikelski2022, Burton2022} and how mito-nuclear incompatibilities can influence the emergence of species, albeit potentially masked by other ecological factors.

Our model's predictions are limited by its simplified nature. Real genetic mechanisms are far more complex. Genes may not express independently, nor are they limited to two alleles as portrayed here, and their mutations rates are influenced by their physical structure.  Additionally, mito-nuclear incompatibilities may have a polygenic genetic architecture that varies among populations \cite{Pereira2021}. The model also does not account co-introgression, where mtDNA invasion is simultaneous or followed by the invasion of co-adapted nuclear genes \cite{Morales2018}, thus minimizing mito-nuclear incompatibilities. Another limitation is that mutations are considered ``neutral'',  neglecting the pervasive purifying selection in mtDNA evolution, which primarily removes non-synonymous mutations \cite{Stewart2008, James2016}. Furthermore, we consider only the `internal' environment, whereas mitochondrial genotypes can also be subject to selection from external factors \cite{Dowling2008, Hill2020b} and respond to environmental cues \cite{Breton2021, Pegan2023}. Future model iterations could explore testing the response with additional selection exclusively on mtDNA, and we posit that fitness gains from the adaptation of the mitochondrial genotype alone might enhance the observed effects.

\section*{Materials and Methods}

Our model, adapted from \cite{aguiar_global_2009,Costa2018,Princepe2022a}, describes a spatially distributed  population evolving under the influence of mito-nuclear selection. Individuals reproduce sexually and are separated into males and females. Each individual is described by its sex, its spatial location in a two-dimensional lattice with periodic boundary conditions, and by two haploid chromosomes representing the nuclear and mitochondrial DNAs. Chromosomes are biallelic and represented by binary strings: for individual $i$, the nDNA is given by $(\sigma_1^i,\sigma_2^i,\dots, \sigma_B^i)$ and the mtDNA by $(\rho_1^i,\rho_2^i,\dots, \rho_{B_m}^i)$, with $B > B_m$ and $\sigma^i_k$, $\rho^i_k$ assuming the values 0 or 1 (Fig. \ref{fig:model}).

Reproduction between individuals $i$ and $j$ of opposite sex is possible only if they are within a maximum spatial distance $R$ from each other and if their nuclear genomes are sufficiently similar, so that $d^n_{ij} = \sum_{k=1}^B |\sigma^i_k-\sigma^j_k| \leq G$ \cite{higgs1991stochastic}. The genetic mating threshold $G$ and the spatial mating neighborhood radius $R$ are parameters of the model. When reproduction is successful, the offspring inherits the nuclear alleles of each parent with equal probability (free recombination) and the full mitochondrial genome from the mother. Alleles are then subjected to mutation with probabilities $\mu$ for nDNA and $\mu_m$ for mtDNA. Mutations can be reversed. The offspring is placed at the location of the reproducing individual (with probability $1-D$) or in one of the 20 closest sites, chosen at random, around the reproducing individual with diffusion probability $D$. 
If the individual cannot find a compatible mate within $R$, it expands its search area to radius $R+1$ and then to $R+2$. If no compatible mate can found in the enlarged area, or if the individual dies without reproducing because of low fitness, another individual is randomly selected within its original mating neighborhood with radius $R$ to reproduce in its place, keeping the community size constant.

Generations are non-overlapping, and each individual has a chance of reproducing proportional to its fitness, $w_i = \exp{(-S d_i^2)}$. Here $d_i=(1/B_m) \sum_{k=1}^{B_m} |\sigma_k^i-\rho_k^i|$ is the normalized mito-nuclear distance for individual $i$ and $S$ is the strength of selection, so that $S=0$ corresponds to the neutral case. Perfectly matched genomes, $d_i=0$, corresponds to maximum fitness. Only the first $B_m$ nuclear alleles interact with the mitochondrial DNA. Initially all nuclear alleles of all individuals are set to 0. For the mitochondrial DNA we set the first $f B_m$ alleles to 1 (``mismatched'') and the remaining $(1-f) B_m$ alleles to zero (``matched'')(Fig. \ref{fig:model}). The strength of selection can also be expressed as $S=1/2\sigma_w^2$, where $\sigma_w$ measures how large the mito-nuclear distance can be without significantly decreasing fitness \cite{Princepe2021,Princepe2022a}.

The probability of an individual reproducing depends on the strength of selection over mito-nuclear compatibility. If selection was absent, the probability that an individual is not selected to reproduce in $N$ trials with replacement is $Q = (1-1/N)^N \approx e^{-1} \approx 0.37$ for large $N$. In the presence of selection, this probability changes to $Q_w(i) = 2Q(w_{max} - w(i))/(w_{max} - w_{min})$, where $w_{max}$ and $w_{min}$ are the maximum and minimum fitness in the population, respectively. Individuals with $w(i) = w_{max}$ have $Q_w(i) = 0$ and probability of reproducing $P_w = 1 - Q_w = 1$. Those with $w(i) = w_{min}$ have $Q_w(i) = 2Q$ and $P_w = 1 - 2Q \approx 0.26$. Therefore, individuals with low fitness still have a small chance of reproducing. A mating partner is selected from a pool of compatible individuals within the mating neighborhood with probability proportional to their fitness. To maintain a constant community size, when an individual dies without reproducing, another one is selected from its mating neighborhood, according to fitness, to reproduce in its place.

We performed simulations considering a standard set of parameters used in previous demonstrations of the model \cite{Princepe2021,Princepe2022a}, ensuring the initiation of radiation and the establishment of equilibrium (when the number of species become constant) within a few hundred generations. 
The parameters values employed were as follows: nuclear and mitochondrial genomes of sizes $B=1500$ and $B_m=500$, respectively; genetic mating threshold of $G=75$, equivalent to $0.05B$; population of $N=1300$ individuals randomly distributed in $L\times L=100\times 100$ living area; nuclear mutation probability of $\mu=2.5 \times 10^{-3}$ per gene; mating neighborhood radius of $R=5$; and a diffusion probability of $D=0.02$. 
We varied the ratio $\mu_m/\mu$ across 3, 5, 8, and 10. In our model, mtDNA features -- finite, asexual, and biallelic -- impose a critical upper limit on the mitochondrial mutation rate, $\mu_c \sim 1/B_m$. Rates exceeding this value disrupt the sequence's information, resembling an error threshold \cite{Princepe2021}. For $B_m=500$, the threshold $\mu_c$ is calculated as $0.002$, equivalent to eight times the nuclear mutation rate.
The simulations spanned a total of $T=2000$ generations.

Previous studies have investigated how variations in these parameters affect the speciation process. In general, increasing $G$ or $R$ promotes gene flow, hindering speciation, while an increase in $B$ or $\mu$ facilitates speciation and reduces the time required to reach equilibrium. Increasing $N$ results in a proportional increase in the number of species \cite{aguiar_global_2009,Costa2018}.
We ran 50 independent simulations for each parameter set and present ensemble-averaged measurements.

We define ``substitutions'' as mutations that become fixed in the population, changing from an initial state of 0 to a final state of 1 or vice versa. Mutations in our model are neutral, meaning they have no impact on individual fitness alone, i.e., all mutations are equivalent. Furthermore, we do not distinguish between synonymous and non-synonymous substitutions; broadly speaking, all substitutions can be interpreted as non-synonymous.

\section*{Acknowledgment}

This work was partly supported by the São Paulo Research Foundation (FAPESP), grants \#2018/11187-8 (DP), and \#2021/14335-0 (ICTP-SAIFR). MAMA was supported by Conselho Nacional de Desenvolvimento Científico e Tecnológico (CNPq), grant \#301082/2019-7. We thank Pedro Cintra for his insightful suggestions on improving the figures presented in this article.

\section*{Author contributions}
 D.P. and M.A.M.A. designed the study; D.P. run the simulations; D.P. and M.A.M.A analyzed the results, wrote the original draft, reviewed and edited the manuscript.

\section*{Conflict of interest}
{The authors declare no conflict of interests.}

\section*{ Code availability }
All Fortran codes used in this study are available in the GitHub repository at https://github.com/deborapr/introgression.

{\footnotesize
\bibliography{referencias}}
\bibliographystyle{naturemag}

\end{document}